\documentclass{amsart}
\usepackage{graphicx}
\usepackage{epsf}
\theoremstyle{definition}

\theoremstyle{remark}

\numberwithin{equation}{section}

\newcommand{\bt}{\begin{tabular}}
\newcommand{\et}{\end{tabular}}
\newcommand{\bi}{\begin{itemize}}
\newcommand{\ei}{\end{itemize}}
\newcommand{\bbm}{\begin{minipage}[h]{6cm} ${}$ \\}
\newcommand{\eem}{\\ ${}$ \end{minipage}}
\newcommand{\bmm}{\begin{minipage}[h]{1.2truecm} ${}$ \\}

\begin{document}

\vspace{-4truecm}
{}\hfill{DSF$-$36/2004}
\vspace{1truecm}

\title[Il corso di Fisica teorica di Ettore Majorana]{Il corso di Fisica teorica di
Ettore Majorana: \\ il ritrovamento del Documento Moreno}%
\author{S. Esposito}%
\address{{\it S. Esposito}: Dipartimento di Scienze Fisiche,
Universit\`a di Napoli ``Federico II'' \& I.N.F.N. Sezione di
Napoli, Complesso Universitario di M. S. Angelo, Via Cinthia,
80126 Napoli ({\rm Salvatore.Esposito@na.infn.it})}%


\begin{abstract}
Viene inquadrato storicamente ed analizzato in dettaglio il corso di Fisica teorica
svolto nel 1938 all'Universit\`a di Napoli da Ettore Majorana, prima della sua misteriosa
scomparsa. In particolare si relaziona sul recente ritrovamento di un quaderno
appartenuto ad Eugenio Moreno in cui sono riportati gli appunti delle lezioni di
Majorana, sei delle quali non sono contenute nella raccolta delle dieci lezioni finora
note e conservate a Pisa.
\end{abstract}
\maketitle


\section{Introduzione}

\noindent L'interesse per l'opera di Ettore Majorana, da parte di chi non lo ha
conosciuto personalmente, pu\`o superficialmente essere generato da un apprezzamento che
ne fece Enrico Fermi e testimoniato da un altro fisico, Giuseppe Cocconi. Poco dopo la
misteriosa scomparsa di Majorana (avvenuta nel marzo del 1938), per far comprendere a
Cocconi che cosa significasse per la Fisica la perdita di Majorana, Fermi si espresse in
tal modo: ``...poi ci sono i geni, come Galileo e Newton. Ebbene, Ettore era uno di
quelli. Majorana aveva quel che nessun altro al mondo ha...'' \cite{Cocconi}. Un giudizio
simile fu espresso da Fermi anche in altre occasioni, pi\`u ufficiali, come ad esempio
nella vicenda del concorso a cattedra del 1937 (che valse a Majorana il posto di
professore ordinario di Fisica teorica ``per alta fama di singolare perizia'',
indipendentemente dalle usuali regole concorsuali), o nella lettera indirizzata a
Mussolini per richiedere ricerche approfondite sulla scomparsa di Majorana \cite{Recami}.
Tali apprezzamenti possono apparire, soprattutto sulle labbra di Fermi, alquanto
esagerati se posti in relazione alla scarna produzione scientifica nota, appena nove
articoli pubblicati su riviste scientifiche, sebbene il nome di Majorana sia largamente
noto tra gli specialisti soprattutto (ma non solo) per l'ipotesi del {\it neutrino di
Majorana}, ampiamente usata dai fisici delle particelle elementari e attualmente al
vaglio dell'osservazione sperimentale.
\\
Una luce sulle singolari capacit\`a di Majorana ci viene invece offerta dalla vasta
produzione di scritti inediti, quasi totalmente conservati alla Domus Galilaeana in Pisa,
una parte dei quali (i ``Volumetti'') \`e stata recentemente pubblicata \cite{volumetti}.
Tuttavia qui ci si occuper\`a del Majorana docente di Fisica teorica, cos\`i come
testimoniato  dai suoi allievi e desunto dalle lezioni tenute all'Universit\`a di Napoli.
Soprattutto si far\`a riferimento all'importante recentissimo ritrovamento di alcuni
documenti contenenti, tra l'altro, sei lezioni precedentemente ignote, a fronte delle
dieci di cui si conserva l'originale manoscritto a Pisa. Il valore di tale ritrovamento,
non solo per il contenuto scientifico dei documenti, ma anche per l'inquadramento storico
particolareggiato delle lezioni e degli ultimi giorni a Napoli di Majorana prima della
sua scomparsa, verr\`a evidenziato qui con un certo dettaglio, avvalendosi anche di altre
testimonianze pi\`u o meno note relative alla figura di docente del fisico catanese. A
tale analisi, per una migliore comprensione, verr\`a premessa una rapida rassegna sui
fatti salienti che portarono Majorana ad insegnare all'Universit\`a di Napoli; ulteriori
interessanti dettagli possono essere trovati in alcune ottime biografie presenti in
letteratura \cite{Recami}, \cite{Amaldi}.

\section{Il concorso a cattedra e l'approdo a Napoli}

\noindent Nel 1937 l'Universit\`a di Palermo, per interessamento di Emilio Segr\`e,
richiese un nuovo concorso per la Fisica teorica. I concorrenti, oltre Majorana (invitato
insistentemente a partecipare al concorso da Fermi e dagli amici), erano Leo Pincherle,
Giulio Racah, Gleb Wataghin, Gian Carlo Wick e Giovanni Gentile (figlio dell' omonimo
filosofo, gi\`a ministro). La commissione giudicatrice, riunitasi a Roma, era presieduta
da Fermi ed era costituita da Antonio Carrelli (segretario), Orazio Lazzarino, Enrico
Persico e Giovanni Polvani.
\\
Prescindendo da diverse possibili interpretazioni, i documenti ufficiali testimoniano che
la commissione prospett\`o al Ministro Bottai (il quale accoglier\`a la proposta)
``l'opportunit\`a di nominare il Majorana professore di Fisica teorica per alta e
meritata fama in una Universit\`a del Regno, indipendentemente dal concorso''
\cite{concorso}. Attribuita ``fuori concorso'' la cattedra a Majorana, la commissione
former\`a poi la terna vincente come segue: 1) G.C. Wick, 2) G. Racah, 3) G. Gentile. Un
membro della commissione, Carrelli, era il direttore dell'Istituto di Fisica
dell'Universit\`a di Napoli, e dovette probabilmente avere un ruolo nella scelta della
sede da assegnare per la cattedra di Majorana. Carrelli, infatti, era sostanzialmente un
fisico sperimentale {\it classico} e conosceva bene l'ambiente napoletano, povero di
fisici teorici {\it moderni}, ma da una lettera di Majorana a Gentile \cite{Gentile} si
apprende che Majorana era ``in rapporti epistolari con Carrelli che \`e veramente una
gran brava persona''.
\\
La nomina a professore ordinario, partecipata dal Ministro Bottai il 2/11/1937, decorse
dal 16 novembre dello stesso anno, ma Majorana si rec\`o a Napoli verso l'inizio
dell'anno successivo (probabilmente il 10/1/1938). Qui si rese subito conto
dell'esiguit\`a del gruppo di fisici napoletani: ``...Praticamente l'Istituto si riduce
alla persona di Carrelli, del vecchio aiuto Maione e del giovane assistente Cennamo. Vi
\`e anche un professore di fisica terrestre difficile a scoprire...'' \footnote{Tale
fantomatico professore era, probabilmente, Giuseppe Imb\'o, Direttore dell'Istituto di
Fisica terrestre e quindi non propriamente appartenente all'Istituto diretto da Carrelli.
\'E anche possibile, per\'o, che si trattasse di qualche libero docente in Fisica
terrestre (come, ad esempio, Giovanni Platania); tuttavia, nessuna notizia certa si
possiede a riguardo.} \cite{MFN1}. Delle perplessit\`a di Majorana per l'ambiente
napoletano (anche il numero degli studenti era molto basso, cinque fisici in tutto, come
si vedr\`a pi\`u avanti) testimoniano allo stesso modo alcuni suoi allievi \cite{Sciuti},
\cite{Senatore}. Tuttavia, in una successiva lettera a Gentile \cite{MGN1}, egli
dichiarer\`a di essere ``contento degli studenti, alcuni dei quali sembrano risoluti a
prendere la fisica sul serio''.

\section{Il corso di Fisica teorica}

\noindent Prima dell'arrivo di Majorana a Napoli, il corso di Fisica teorica era tenuto
dal direttore Carrelli, e gli argomenti trattati non riguardavano affatto i moderni
sviluppi della Fisica quantistica (Sebastiano Sciuti riferisce ironicamente, ad esempio,
che un argomento {\it avanzato} del corso di Carrelli era quello dei moti browniani...
\cite{Sciuti}).

\subsection{La lezione inaugurale}

\noindent Majorana annuncia l'inizio del suo corso per il gioved\`i 13 gennaio alle nove,
ma concorda con il preside della Facolt\`a ``di evitare ogni carattere ufficiale
all'apertura del corso'' \cite{MFN1}. Di ci\`o se ne ha traccia indiretta nell'assenza di
una tale notizia sui giornali cittadini (come, ad esempio, {\it Il Mattino}),
contrariamente a quanto avveniva per altri corsi, certamente pi\`u affollati.
\\
Secondo la testimonianza di Gilda Senatore \cite{Senatore}, alla lezione inaugurale non
parteciparono gli studenti del corso medesimo (o per espressa indicazione del direttore o
per cause contingenti). Lo stesso Majorana afferma che ``non \`e stato possibile
verificare se vi sono sovrapposizioni d'orario, cos\`i che \`e possibile che gli studenti
non vengano e che si debba rimandare'' \cite{MFN1}. Di fatto, la prolusione si terr\`a
come previsto il 13 gennaio nell'aula grande di Fisica sperimentale in Via Tari, senza
gli studenti ma con una decina di partecipanti, secondo quanto ricorda Sciuti
\cite{Sciuti}, forse anch'egli presente alla lezione inaugurale. \`E da notare che se \`e
fondata l'ipotesi secondo la quale la non partecipazione degli studenti fu espressamente
voluta (da Carrelli o da altri), ci\`o potrebbe trovare una valida spiegazione in una
antica consuetudine dell'Universit\`a di Napoli secondo cui il nuovo docente doveva {\it
dimostrare} agli altri professori dell'Universit\`a di essere meritevole del posto che
sarebbe andato ad occupare \cite{Preziosi}. Tale consuetudine sarebbe rimasta in vigore
fino all'insediamento di Felice Ippolito alla Facolt\`a di Ingegneria nell'immediato
dopoguerra. Tuttavia, l'accenno riportato sopra ad ``evitare ogni carattere ufficiale
all'apertura del corso'' e la presenza dei familiari di Majorana alla prolusione
\cite{Recami}, renderebbero solo probabile il perpetuarsi dell'antica tradizione
fridericiana nel caso di Majorana.
\\
La successiva lezione, l'inizio vero e proprio del corso, si tenne il sabato 15 gennaio e
il corso proseguir\`a fino a marzo nei giorni pari (marted\`i, gioved\`i e sabato). La
sede delle lezioni era un'auletta posta di fronte a quella usata per la prolusione,
l'``aula di Fisica superiore'' e l'``aula di Fisica teorica'', situata al pian terreno
dell'Istituto di Via Tari e che affacciava su un largo all'interno del cortile
dell'Universit\`a \cite{Sciuti}.

\subsection{Gli studenti del corso}

\noindent Gli studenti {\it fisici} che partecipavano al corso erano cinque: Nella
Altieri, Laura Mercogliano, Nada Minghetti, Gilda Senatore e Sebastiano Sciuti. Le
quattro studentesse erano tutte ``allievi interni'' (per cui oltre a studiare per i
corsi, svolgevano anche attivit\`a di ricerca, prevalentemente in Fisica classica), e
seguirono le lezioni di Majorana quando erano al loro quarto anno di Fisica
\cite{Senatore}; una delle loro maggiori preoccupazioni era quella di superare esami,
essendosi trovate ``fuori corso'' \footnote{Si osservi che, nel corso di Laurea in Fisica
negli anni considerati, il corso di Fisica teorica si svolgeva in due parti (come anche
altri corsi) tenute, rispettivamente, al terzo ed al quarto anno. Ad eccezione della
Senatore, che si laure\'o un anno pi\'u tardi, gli altri quattro studenti si laurearono
invece nel dicembre del 1938.} \cite{Sciuti}. Sciuti, invece, aveva gi\`a seguito un
corso di Fisica teorica tenuto da Carrelli. Tuttavia egli era desideroso di entrare in
contatto con il gruppo di Roma guidato da Fermi (si era iscritto a Fisica a 17 anni
proprio accogliendo l'invito di Orso Mario Corbino e Fermi a Roma), e seguire il corso di
Majorana (che proveniva dal gruppo di Roma) appariva ai suoi occhi come il primo passo
nel raggiungimento del suo obiettivo. Nella loro attivit\'a di ricerca, tutti gli allievi
si occupavano prevalentemente di Fisica sperimentale \footnote{Gli argomenti specifici
delle ricerche possono essere desunti dai titoli delle loro Tesi di Laurea: {\it
Sull'emissione spettrale in ultrarosso di alcuni fosfati e silicati} (Altieri), {\it
Bande d'assorbimenti d'infrarosso dovute alla presenza del gruppo OH} (Mercogliano), {\it
Emissione totale di metalli al di sotto e al di sopra del punto di fusione} (Minghetti),
{\it Assorbimento e fluorescenza dei sali di chinino} (Senatore), {\it Ricerche
preliminari nell'ultarosso} (Sciuti).}; ad esempio la Altieri veniva seguita
dall'assistente Cennamo, mentre la Senatore, sebbene incline alla Fisica teorica, si
occupava di Fisica molecolare con l'aiuto Maione e, dopo la scomparsa di Majorana, con
Cennamo \cite{Senatore}.
\\
Oltre ai cinque studenti di Fisica ricordati sopra, vanno poi aggiunti altri uditori
pi\`u o meno assidui del corso di Fisica teorica: Mario Cutolo \cite{Senatore},
\cite{Sciuti} e don Savino Coronato \cite{Senatore}. Il primo, riferisce la Senatore
\cite{Senatore}, probabilmente partecipava al corso anche perch\`e invaghito di un'altra
studentessa, Nada Minghetti, mentre il secondo era uno studente di Matematica (che si
laureer\'a con Renato Caccioppoli sempre nel 1938), e diventer\'a poi l'assistente fedele
di Caccioppoli all'Istituto di Analisi Matematica; fu probabilmente invitato a seguire il
corso dallo stesso Caccioppoli, che aveva partecipato alla lezione inaugurale. In
aggiunta a queste informazioni gi\'a note, si \`e poi recentemente scoperto \cite{Moreno}
che, oltre ai due uditori appena menzionati, si deve tener conto di un altro probabile
assiduo uditore del corso: Eugenio Moreno. Tale presenza al corso di Majorana non era
(fino ad oggi) n\'e sospettata n\'e tantomeno documentata, in quanto i testimoni viventi
del corso (sostanzialmente Sciuti e la Senatore) non ne avevano mai accennato. La notizia
di tale presenza \`e invece emersa solo recentemente dal figlio di Moreno (deceduto nel
2000), Cesare \cite{Moreno}, ed \`e avvalorata dal ritrovamento degli appunti delle
lezioni ricopiati da Eugenio Moreno dagli originali di Majorana.
\\
Eugenio Moreno nacque a Napoli nel 1910 e si iscrisse al corso di Laurea in Matematica
nel 1929-1930. Dopo diversi periodi caratterizzati da rinvio militare intervallati da
altri periodi di frequenza di corsi di Leva militare, il 30/12/1937 viene messo in
congedo (ma successivamente verr\'a richiamato alle armi) quale Sottotenente di
Complemento. Il corso di Majorana, diversamente da altri corsi della Facolt\'a di Scienze
che iniziavano usualmente a novembre (inizio dell'anno accademico) anzich\`e a gennaio,
parte dopo pochi giorni del rientro di Moreno all'universit\'a.

\subsection{Lo stile}

\noindent Majorana, ``vestito di blu''\cite{Sciuti}, aveva sempre un aspetto ``triste e
perplesso'' e ci\`o, unito alla non facile comprensione degli argomenti avanzati che egli
trattava a lezione \footnote{La Senatore ricorda, tra l'altro, come la matematica che
egli usava non era affatto introdotta negli altri corsi universitari \cite{Senatore}.}
certamente infondeva una certa soggezzione nei giovani uditori del corso di Fisica
teorica. D'altro canto, il soprannome di {\it Grande Inquisitore} gli era stato
attribuito gi\`a molti anni prima dagli amici di sempre del gruppo di Fermi a Roma
\cite{Amaldi}, \cite{Recami}. E anche al di fuori del contesto ``ufficiale'' delle
lezioni, Majorana confermava questo comportamento: ``salutava e rispondeva gentilmente al
saluto e, magari, timidamente sorrideva; si intuiva che doveva essere profondamente buono
e sensibilissimo, ma non fu mai estroverso o invitante, anzi fu sempre estremamente
schivo'' \cite{Sena98}. E ancora: ``in quel lungo corridoio buio al piano terra...
camminava sempre rasente al muro, silenziosamente e solo, muovendosi come un'ombra''
\cite{Sena98}.
\\
Quando arriv\`o a Napoli, certamente il direttore Carrelli dovette parlare degli studenti
e delle loro ricerche a Majorana \cite{Senatore}, il quale si dovette rendere subito
conto del singolare compito che si era apprestato ad accettare con un cos\'i esiguo
numero di studenti \cite{Senatore}, come gi\`a accennato sopra. Tuttavia egli era
fermamente deciso a portare a termine in maniera responsabile il compito assunto
\cite{Sena98}.
\\
A lezione ``era chiarissimo nella trattazione dell'argomento che proponeva di volta in
volta all'inizio della lezione e che svolgeva con dovizia di particolari, dando sempre la
prevalenza alla parte fisica pi\`u che a quella matematica; ma quando si volgeva alla
lavagna e cominciava a scrivere, faceva calcoli che sul momento non sempre si riusciva a
seguire'' \cite{Sena98}. Il carattere di Majorana, poi, certamente non invitava i timidi
studenti a interromperlo per chiedergli spiegazioni. Talvolta alcune domande gli venivano
esplicitamente rivolte solo da Sciuti \cite{Sciuti}, il quale gli chiese anche se poteva
avvalersi dell'ausilio di qualche testo nel seguire le lezioni del corso. A ci\`o
Majorana rispose che avrebbe distribuito degli appunti, e che comunque avrebbe seguito il
recente testo di Persico \cite{Persico} (``un libro in italiano, molto bello''), sebbene
apportando alcune ``semplificazioni formali'' \cite{Sciuti}. Sulla base degli appunti
delle lezioni pervenuti sino a noi (vedi il prossimo paragrafo), possiamo tuttavia
concludere che tali ``semplificazioni'' non erano affatto sporadiche e accessorie, e che
quindi l'impostazione di Majorana era ben diversa da quella usata da Persico
\cite{Drago}. Un altro testo ``consigliato'' a Sciuti fu quello di Heisenberg
\cite{Heisenberg}, ma anche per questo vale la considerazione svolta sopra \footnote{\`E
da segnalare che entrambi i testi citati facevano parte della scarna (appena una
quindicina di testi) biblioteca personale di Majorana; in particolare quello di
Heisenberg era posseduto nella sua versione originale tedesca del 1930 \cite{EMJR}.
Inoltre il riferimento bibliografico al testo di Persico sembra apparire (sebbene
cancellato dall'usura del tempo) sul retro del Documento Moreno (sovraccoperta
cartacea).}
\\
Gli studenti in difficolt\`a con la comprensione degli argomenti
trattati non potevano, quindi, contare sull'aiuto di un libro di
testo, e la Senatore ricorda che ``solo gli appunti presi durante
le lezioni e raccordati tra noi dopo, ci permettevano di correlare
la parte teorica, magistralmente spiegata, con quella matematica
che la giustificava'' \cite{Sena98}. Infatti, gli studenti
(probabilmente solo quelli ``fisici'') usualmente si incontravano
il giorno successivo a quello del corso per ``confrontare'' gli
appunti presi a lezione e studiare insieme gli argomenti relativi.
Talvolta a lezione, quando Majorana si accorgeva (interrompendosi
e voltandosi indietro) che gli studenti stentavano a capire ci\`o
che lui stava esponendo, si fermava e rispiegava lo stesso
argomento \cite{Senatore}. ``Proprio durante qualcuna di quelle
lezioni pi\`u aride e pi\`u pesanti in quanto l'argomento trattato
era afferente essenzialmente a metodi matematici da applicarsi
allo studio di fenomeni fisici, Majorana dimenticava forse di
essere quel grandissimo scienziato che era, perch\`e mentre era
alla lavagna e scriveva, improvvisamente si fermava, poi si
volgeva, ci guardava un attimo, sorrideva e riproponeva la
spiegazione facendo aderire il concetto gi\`a esposto alle formule
che riempivano la lavagna'' \cite{Sena98}.

\section{Le lezioni all'Universit\`a}

\noindent Come visto sopra, secondo quanto ricordato da Sciuti \cite{Sciuti}, Majorana
appront\`o degli appunti delle sue lezioni da dare ai suoi studenti per facilitarli nella
comprensione degli argomenti svolti. Probabilmente (vedi pi\`u avanti) ci\`o avvenne dopo
il 22 gennaio 1938, ossia dopo la sua quinta lezione, in quanto in tutti i documenti
attualmente noti mancano proprio le prime cinque lezioni (o, pi\`u precisamente, le prime
quattro lezioni, pi\'u la prolusione al corso). La storia di come tali appunti,
importantissimi per comprendere l'opera innovatrice di Majorana quale docente, sono
giunti ai nostri giorni \`e piuttosto intrigante, e vale quindi la pena analizzarla con
l'ausilio di tutti i documenti a disposizione.

\subsection{Il ritrovamento del Documento Moreno}

\noindent Nel settembre 2004, A. Drago e chi scrive hanno raccolto la preziosa
testimonianza del figlio di Eugenio Moreno, Cesare, secondo il quale il padre sarebbe
stato uno degli assidui frequentatori (non fisici) del corso di Majorana. In tale
occasione si \`e venuti a conoscenza dell'esistenza di un importante documento (che qui
verr\`a chiamato, brevemente, Documento Moreno), in cui Moreno stesso ricopi\`o
fedelmente gli appunti delle lezioni di Majorana.
\\
Prescindendo da tale documento, fino ad ora gli interessati avevano a disposizione solo
alcuni originali manoscritti di Majorana, conservati alla Domus Galilaeana a Pisa, e
pubblicati alcuni anni or sono in stampa anastatica \cite{Bibliopolis}. L'analisi di tali
originali aveva portato a concludere \cite{Cabibbo} che, eccezion fatta al pi\`u per una
o due lezioni ritenute mancanti, questi coprissero sostanzialmente tutto il corso di
Fisica teorica. Alcuni dubbi \cite{Sena98} su tale conclusione, sostanzialmente basati su
ricordi lontani, erano stati sollevati proprio da una studentessa del corso, G. Senatore,
la quale ha sempre concordemente affermato che ``mancano alcuni capitoli delle lezioni,
il cui testo mi fu consegnato, completo persino degli appunti di quella che il professore
avrebbe dovuto tenere il giorno successivo alla sua scomparsa. Manca ancora un esiguo
gruppo di fogli, scritti anche essi in originale ed in maniera ordinata come gli altri,
ma non facenti parte delle lezioni gi\`a tenute'' \cite{Sena98}.
\\
L'importanza notevole del Documento Moreno \`e duplice: da un lato esso \`e giunto fino a
noi in maniera {\it completamente} indipendente dal percorso seguito dai manoscritti
originali (vedi pi\`u avanti), dall'altro {\it tutte} le lezioni del corso vengono in
esso {\it numerate} e {\it datate}, contrariamente a ci\`o che avviene per i manoscritti
della Domus Galilaeana (alcune lezioni (le prime) sono datate, mentre le altre (dalla N.
15) sono numerate). Ci\`o permette, quindi, di ricostruire in maniera dettagliata e
univoca tutta l'evoluzione del corso di Fisica teorica tenuto e, forse, pu\`o tornare
utile anche ai fini dell'indagine storica sugli ultimi giorni di Majorana a Napoli prima
della sua scomparsa.
\\
Soffermandosi, poi, anche in maniera superficiale sul testo contenuto nel Documento
Moreno, e confrontandolo con quello dei manoscritti conservati a Pisa \cite{Bibliopolis},
risultano subito evidenti due grosse peculiarit\`a: 1) il Documento Moreno presenta {\it
sei lezioni inedite} non contenute nei documenti della Domus Galilaeana; 2) le lezioni
comuni ad entrambi gli archivi sono {\it completamente identiche}, per cui la copia
effettuata da Moreno deve ritenersi del tutto fedele agli originali. Quest'ultima
conclusione \`e confermata, per quanto riguarda il contenuto scientifico e il modo di
presentarlo, dall'analisi approfondita delle sei lezioni inedite e dal confronto con
altri scritti di Majorana (i cinque ``Volumetti'' \cite{volumetti} e i diciotto
``Quaderni'' conservati a Pisa).

\begin{figure}
\begin{center}
\vspace{1.5truecm} \epsfysize=17.6cm \epsfxsize=12.5cm \epsffile{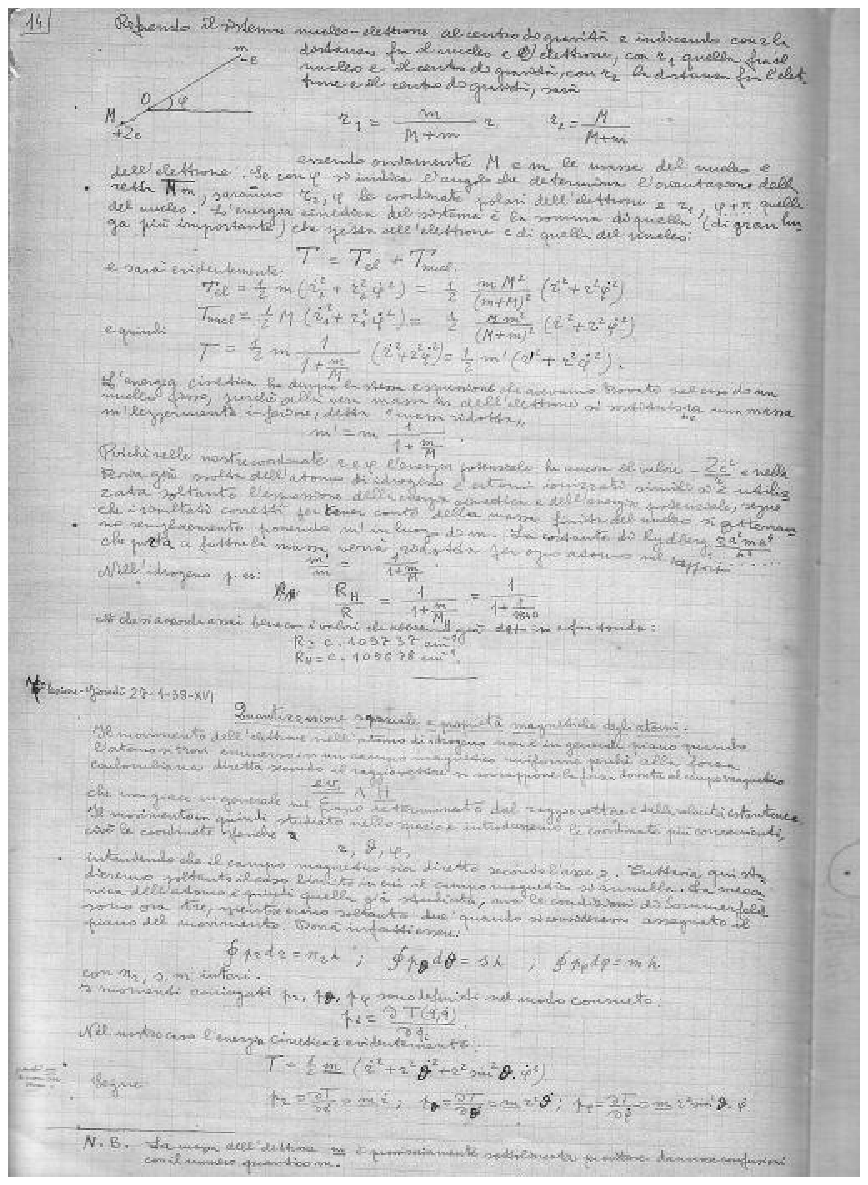}
\caption{Una pagina del Documento Moreno; qui inizia la lezione N.7, la prima lezione
inedita di Majorana (la parte iniziale della pagina \`e invece presente negli originali
conservati a Pisa; cfr. \cite{Bibliopolis}). Si osservi come Eugenio Moreno, l'autore del
documento, abbia successivamente corretto la numerazione della lezione o come abbia
aggiunto una piccola chiosa in basso a sinistra.} \label{fig1}
\end{center}
\end{figure}

\subsection{Dal 1938 ad oggi}

\noindent Gli appunti delle lezioni di Majorana ci sono giunte attraverso diversi
percorsi, non tutti facilmente identificabili, e alcuni punti oscuri permangono tuttora.
\\
Il giorno 25 marzo, venerd\`i, non era prevista alcuna lezione di Fisica teorica (il
giorno prima Majorana aveva tenuto la sua ventunesima lezione) ma, secondo quanto
racconta la Senatore, ``Majorana contrariamente a quanto di solito faceva, venne in
Istituto e vi si trattenne soltanto pochi minuti'' \cite{Sena98}. Come ricordato sopra, i
giorni seguenti alle lezioni, gli studenti si riunivano per studiare gli argomenti
trattati in aula, ed \`e probabile che Majorana era a conoscenza di ci\`o, per cui venne
appositamente in Istituto quel venerd\`i di marzo per consegnare i suoi appunti
\cite{Senatore}. ``Dal corridoio che immetteva nell'auletta in cui mi trattenevo
scrivendo, mi chiam\`o per nome: `Signorina Senatore...'; non entr\`o ma si trattenne nel
corridoio; lo raggiunsi ed egli mi consegn\`o una cartella chiusa dicendomi: `ecco,
prenda queste carte, questi appunti... poi ne riparleremo'; poi and\`o via e voltandosi
ripet\`e: `poi ne riparleremo' '' \cite{Sena98}. Dopo quel venerd\`i 25 marzo, la
Senatore non si rec\`o all'Istituto di Fisica (dalla provincia di Salerno dove risiedeva)
per circa 15 giorni per motivi di salute, e le carte furono conservate nella sua
abitazione per molto tempo \cite{Senatore}. La Senatore tenne il riserbo su questa
faccenda per molti mesi, ma intanto, dalla testimonianza di un'altra studentessa, N.
Minghetti, si \`e venuti a conoscenza del fatto che alcuni giorni dopo la scomparsa di
Majorana, Carrelli aveva chiesto agli studenti gli appunti presi durante le lezioni
\cite{Russo}. Si intendeva qui, probabilmente, la copia degli appunti di Majorana fatta
dagli studenti o gli appunti presi direttamente dagli studenti a lezione, ma in ogni caso
\`e rimarchevole l'interesse verso questi appunti gi\`a qualche giorno dopo la scomparsa
di Majorana.
\\
Allorquando, verso la fine del 1938, la Senatore entr\`o in stretti
rapporti con Francesco Cennamo, assistente del direttore Carrelli, la giovane studentessa
fece visionare gli appunti di Majorana da Cennamo, il quale, dopo qualche tempo, ad
insaputa della Senatore li port\`o a Carrelli, forse per discutere di qualche argomento
specifico \cite{Senatore}. Carrelli, essendo il consegnatario ufficiale di tutti gli
effetti di Majorana, ritenne opportuno di non riconsegnare a Cennamo gli appunti in
questione.
\\
La storia successiva degli appunti diventa molto fumosa. Da una lettera del
2/12/1964 di Gilberto Bernardini ad Edoardo Amaldi, che gli aveva chiesto notizie sugli
appunti, si apprende quanto segue: ``queste sono le pagine delle lezioni di Majorana che
finalmente, nel mio sgombero da Ginevra, ho ritrovato. Sono solo una parte. Le altre mi
sembra che le avesse Giovannino Gentile o forse non furono mai scritte; precisamente le
linee confuse di un ricordo troppo lontano, mi fanno credere che me le avesse date
Giovannino per aiutarmi a capire il Dirac'' \cite{Bernardini}. A far data dal 1965,
dunque, gli appunti di Majorana sono nelle mani di Amaldi, il quale successivamente li
depositer\`a, insieme con altri scritti, alla Domus Galilaeana. Tuttavia sul passaggio
Carrelli-Bernardini (o Carrelli-Amaldi, supponendo che prima di Bernardini li avesse
gi\`a avuti Amaldi) non esiste attualmente alcuna documentazione probante. Sembra,
comunque, un dato acquisito \cite{Senatore} il fatto che la Senatore venne di nuovo a
conoscenza dell'esistenza degli appunti di Majorana solo dopo la morte del marito
Cennamo, quando un altro ex-assistente di Carrelli, Elio Tartaglione, su mandato di
Cennamo ancora vivente, le rivel\`o come avvenne il passaggio degli appunti a Carrelli
(come menzionato sopra) e consegn\`o il libro con la stampa anastatica delle lezioni
\cite{Bibliopolis}. Sul passaggio delle carte di Majorana da ``Napoli'' a ``Roma'' si
possono, dunque, avanzare solo congetture, che per\`o qui non verranno prese in
considerazione. Deve, tuttavia, essere tenuto presente che se Bernardini le avesse
ricevute da G. Gentile \cite{Bernardini}, gli appunti delle lezioni dovrebbero essere
giunti (integralmente o in parte) al gruppo di Fisica di Roma prima del 1942, anno della
morte prematura di Gentile, sincero amico di Majorana e figlio del noto senatore,
conosciuto da Carrelli.
\\
Il tragitto temporale del Documento Moreno \`e invece molto
pi\`u lineare: la copia degli appunti di Majorana \`e sempre stata conservata da Moreno,
e recentemente recuperata dal figlio nell'abitazione paterna \cite{Moreno}.
\\
Un successivo approfondito confronto tra i due documenti ora disponibili potr\`a, quindi,
gettare nuova luce sulla parte mancante dei manoscritti originali.

\subsection{Caratteristiche delle lezioni presenti nel Documento Moreno}

\noindent La numerazione e la datazione di ciascuna delle lezioni svolte da Majorana a
Napoli tra il gennaio e il marzo del 1938, presente nel Documento Moreno, permette una
ricostruzione sufficientemente fedele del corso di Fisica teorica svolto dal grande
scienziato. Va tuttavia tenuto presente che per molte informazioni esiste un'unica fonte
disponibile, per cui la discussione presentata nel seguito pu\`o non essere scevra da
possibili difficolt\'a interpretative. Nella Tabella 1 viene ricostruito un indice delle
lezioni: la numerazione, la data di svolgimento e l'argomento trattato nelle lezioni
vengono riportati cos\`i come compaiono nel Documento Moreno (eccetto per l'ultima riga
relativa a materiale presente solo alla Domus Galilaeana). \`E da notare che, ad
eccezione della lezione N.1, la numerazione delle lezioni fatta da Moreno presenta delle
cancellature fino alla lezione N.16: un dato numero scritto viene successivamente
cancellato e sostituito con quello seguente. Risulta quindi evidente che, mentre Majorana
considerava la prolusione al corso come la sua prima lezione, Moreno cominci\`o a
numerare le lezioni iniziando da quella successiva alla prolusione, a partire dalla quale
erano presenti gli studenti. Tale ``errore'' non viene pi\`u commesso a partire dalla
lezione N.17: considerando che nei manoscritti originali di Majorana la numerazione \`e
presente solo dalla N.15 in avanti (e riportata, in aggiunta, anche nel Documento
Moreno), si pu\`o concludere che i manoscritti delle lezioni N.15 e N.16 (e forse anche
le precedenti) vennero in possesso di Moreno non prima della pausa per le festivit\`a di
Carnevale (vedi pi\`u avanti), ossia dopo l'8 marzo.
\\
La lezione inaugurale fu tenuta, come noto, il 13 gennaio, ma il testo corrispondente non
\`e presente n\'e nel Documento Moreno n\'e alla Domus Galilaeana, e fu ritrovato solo
nel 1972 \cite{Recami} da Erasmo Recami tra le carte conservate dalla famiglia Majorana.
Nel Documento Moreno \`e presente la frase: ``Introduzione: argomenti che saranno
trattati nel corso'', forse indicante che Moreno era presente a tale lezione.
\\
Le successive quattro lezioni, svolte dal sabato 15 al sabato 22 gennaio, non sono
pervenute attraverso nessun documento noto. Solamente per queste lezioni, Moreno lascia
qualche pagina bianca non scritta, aspettandosi probabilmente di ricevere, anche per
queste, gli appunti di Majorana. Gli argomenti qui trattati risultano, quindi, ignoti ma,
seguendo Cabibbo \cite{Cabibbo}, si pu\`o ritenere che Majorana avesse introdotto la
fenomenologia pi\`u rilevante per la Fisica atomica. Tale prima parte del corso si
estender\`a fino alla pausa per le festivit\`a di Carnevale, e quindi fino alla lezione
N.15 del 17 febbraio.
\\
Il testo delle lezioni
dalla N.6 alla N.21 \`e integralmente presente nel Documento Moreno, e per l'analisi del
contenuto delle lezioni conservate anche alla Domus Galilaeana si rimanda alla
discussione in \cite{Cabibbo}.
\\
La lezione N.7 \`e la prima presente nel Documento Moreno ma mancante nell'archivio di
Pisa. In essa viene affrontata la quantizzazione del momento angolare e viene introdotta
l'ipotesi dell {\it spin}, che servir\`a poi di base per le successive lezioni sul
sistema periodico degli elementi e sugli spettri atomici. L'esistenza di tale lezione,
niente affatto sospettata prima del rinvenimento del Documento Moreno, conferma quindi
l'attitudine di Majorana di introdurre in modo dettagliato ed esauriente i concetti nuovi
che gli studenti non avevano affrontato in corsi precedenti.
\\
La lezione di marted\`i 1 febbraio non viene svolta, probabilmente per il concomitante XV
anniversario della fondazione della Milizia Fascista.
\\
La lezione N.10 del 5 febbraio \`e la seconda lezione non presente alla Domus Galilaeana
ma riportata nel Documento Moreno. In essa si prosegue con la teoria classica
dell'irraggiamento (un argomento su cui Majorana si \`e spesso soffermato nei suoi studi
personali \cite{volumetti}), e viene anche discusso un peculiare (e semplice) fenomeno
fisico (la diffusione della luce solare da parte dell'atmosfera) difficilmente (e
stranamente) presente in un corso di Fisica teorica. Sull'esistenza di tale lezione (o
almeno per la sua prima parte) gi\`a si era ipotizzato \cite{Cabibbo}, in quanto la
lezione precedente (la N.9) era evidentemente incompleta.
\\
Le successive quattro lezioni, svolte da marted\`i 8 a marted\`i 15 febbraio, anch'esse
non presenti alla Domus Galilaeana, sono invece completamente inaspettate, in quanto
trattano di un argomento di cui non si sospettava affatto: la Teoria della Relativit\`a.
Tale argomento, importantissimo per la Fisica, nel 1938 ancora era sporadicamente
avversato da alcuni fisici in Italia (si ricordi, ad esempio, l'opposizione di Quirino
Majorana, zio di Ettore), e non era usualmente trattato nei corsi di Laurea in Fisica.
Dal Documento Moreno si apprende, quindi, che fu Majorana il primo fisico ad introdurre
in un corso di Laurea in Fisica a Napoli la Teoria della Relativit\`a \footnote{Si noti,
tuttavia, che questa (soprattutto la Teoria della Relativit\'a Generale), era in quegli
anni materia affrontata dai matematici piuttosto che dai fisici, come ad esempio il caso
dell'insigne matematico Roberto Marcolongo, gi\'a docente all'Universit\'a di Napoli.}.
In queste quattro lezioni, egli la espone secondo il suo abituale modo di procedere,
partendo dalla semplice fenomenologia e introducendo solo dopo il formalismo matematico.
Nella prima lezione di questo ``inserto'', Majorana dapprima discute il principio di
relativit\`a galileiana, poi passa alla questione dell'esistenza dell'etere con
l'esperimento di Michelson e Morley, ed infine introduce (in un modo interessante e
semplice) le trasformazioni di Lorentz, con applicazione al caso del campo
elettromagnetico. Nella successiva lezione discute, quindi, gli aspetti formali della
Teoria della Relativit\`a di Einstein, e ritorna poi nuovamente a considerare il caso
elettromagnetico (ma questa volta soffermandosi sui potenziali invece che sui campi). La
questione della somma relativistica delle velocit\`a viene invece affrontata (partendo,
in modo originale, dalle formula di Fresnel per l'esperimento di Fizeau sulla misura
della velocit\`a della luce) nella terza lezione in oggetto, in cui viene anche discussa
l'invarianza relativistica della carica elettrica. Tra la fine della terza e l'inizio
della quarta lezione  dell'``inserto di Relativit\`a'' \`e poi introdotta in modo molto
dettagliato, e partendo da un principio variazionale (vedi anche \cite{volumetti}), la
dinamica relativistica dell'elettrone. Le lezioni mancanti nell'archivio di Pisa
terminano, quindi, con la discussione dell'effetto fotoelettrico (e la sua
interpretazione einsteiniana) e della diffusione Thomson. Proprio guardando a questi
ultimi due argomenti e al seguente nella successiva lezione (l'effetto Compton), si pu\`o
allora ipotizzare che Majorana abbia introdotto la Teoria della Relativit\`a nel suo
corso per permettere una discussione approfondita e chiara di quei fenomeni appena citati
che, pur facendo parte del grande alveo della Fisica quantistica (argomento di elezione
dei corsi di Fisica teorica moderni), richiedevano l'uso della Teoria della Relativit\`a
per una loro corretta interpretazione.
\\
Il corso si interrompe gioved\`i 17 febbraio per poi riprendere marted\`i 8 marzo con
l'introduzione del formalismo della Meccanica Quantistica. Tale lunga interruzione \`e
certamente giustificata, almeno in parte, dalle festivit\`a per il Carnevale (i
corrispondenti giorni di vacanza accademica andavano dal 24 febbraio al 2 marzo);
tuttavia, il forse eccessivo periodo di vacanza potrebbe anche essere correlato ad altri
avvenimenti di rilievo (appresi dai quotidiani dell'epoca) quali, ad esempio, lo sbarco a
Napoli di Bruno Mussolini e dei suoi ``sorci verdi'' il 22 febbraio con la concomitante
adunata dei Fasci Universitari, e la morte di G. D'Annunzio il 2 marzo.
\\
Il testo degli appunti delle lezioni dalla N.15 all'ultima N.21 del 24 marzo riportato
nel Documento Moreno \`e identico a quello presente nel manoscritti della Domus
Galilaeana, ed \`e stato gi\`a discusso in \cite{Cabibbo}. Vi \`e solo da notare, da un
punto di vista storico, lo slittamento della lezione prevista per marted\`i 15 marzo
``per la venuta di S.M. il Re'' (come riportato nel Documento Moreno) \footnote{Il Re
Vittorio Emanuele III, con il Ministro dell'Educazione Bottai, inaugur\`o a Napoli in
tale data la ``Mostra sui Tre Secoli di Pittura'' napoletana dal 1600 al 1800.} e la
lezione di sabato 19 marzo, soppressa per la ``festa di S. Giuseppe'' (vedi Documento
Moreno).
\\
I documenti relativi alle lezioni di Majorana conservati alla Domus Galilaeana terminano
con un appunto non numerato e non datato, sostanzialmente diverso dai precedenti (con
molte cancellature, sembra in forma preliminare \cite{Bibliopolis}), che non \`e presente
nel Documento Moreno. La Senatore, studentessa del corso, interpreta (\cite{Sena98}, vedi
anche \cite{Cabibbo}) come la lezione che ``il professore avrebbe dovuto tenere il giorno
successivo alla sua scomparsa''. Tale ipotesi \`e certamente molto verosimile, ma \`e
comunque difficile non notare che gli argomenti qui trattati in maniera molto estesa
(sostanzialmente si tratta di applicazioni di Fisica atomica e molecolare) sembrano non
collegati direttamente a quelli della lezione N.21.

\begin{figure}
\begin{center}
\vspace{1.5truecm}
\epsfysize=17.6cm \epsfxsize=12.5cm \epsffile{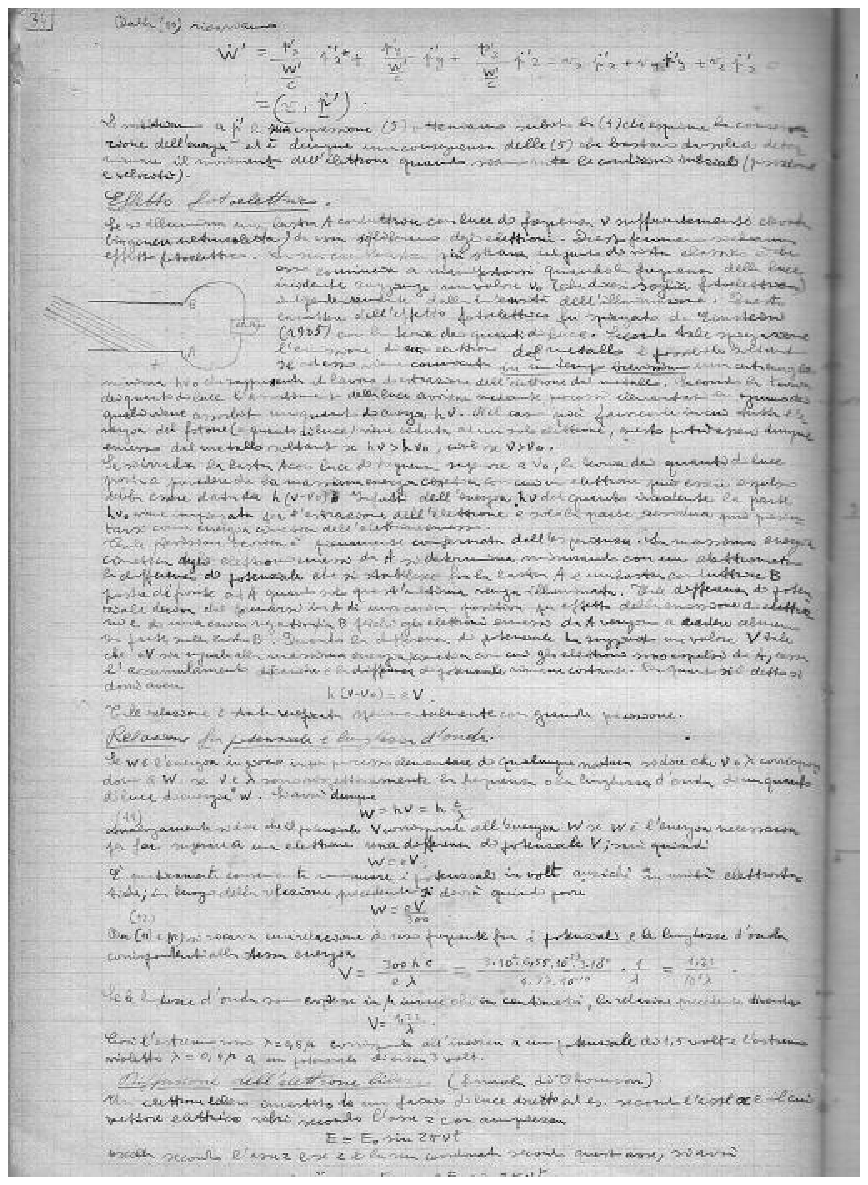}
\caption{Un'altra pagina del Documento Moreno; qui Majorana termina la lezione N.14
(l'ultima inedita) con l'applicazione della Teoria della Relativit\'a all' Effetto
Fotoelettrico. La successiva lezione N.15, presente anche negli originali conservati a
Pisa \cite{Bibliopolis}, continuer\'a con la discussione degli effetti quantistici la cui
spiegazione richiede anche l'uso della Relativit\'a, come ad esempio l'effetto Compton.}
\label{fig2}
\end{center}
\end{figure}

\section{Conclusione}

\noindent Il ritrovamento del Documento Moreno, unitamente ad alcune testimonianze finora
poco o per nulla conosciute, permettono un inquadramento storico molto dettagliato del
corso di Fisica teorica svolto a Napoli da Majorana.

L'importanza di tale ritrovamento risiede innanzitutto nel fatto che le 53 pagine
manoscritte da Eugenio Moreno riportano, senza ombra di dubbio, il testo degli appunti
delle lezioni approntati da Majorana per gli studenti. Tale conclusione emerge
chiaramente da molteplici confronti del testo, presente nel Documento Moreno, con quello
delle lezioni autografe originali conservate a Pisa, cos\'i come con quello presente in
molte altre pagine (edite ed inedite) scritte da Majorana (principalmente i cinque
``Volumetti'' e i diciotto ``Quaderni'', anch'essi conservati a Pisa). Da un punto di
vista strettamente scientifico, l'attuale ritrovamento ha portato alla scoperta di sei
lezioni non note in precedenza, quattro delle quali hanno per argomento la Teoria della
Relativit\`a, inaspettatamente presente nel corso di Fisica teorica di Majorana.
\\
In aggiunta al testo ricopiato dagli originali di Majorana, il Documento Moreno riporta
anche delle significative annotazioni, non presenti negli originali stessi (e non
desumibili da altre fonti, se non in maniera parziale e approssimata) e che consistono,
principalmente \footnote{Nel Documento sono anche riportate alcune chiose a margine del
testo, scritte da Moreno e volte, tipicamente, a segnalare o un abuso di notazione, o la
numerazione impropria di qualche equazione o, ancora, alcune parole scritte da Majorana e
non comprese da Moreno. Tali annotazioni sono rilevanti soprattutto per un'analisi
approfondita del testo e dei contenuti del Documento Moreno che, per\'o, non \`e il fine
del presente lavoro.}, nella numerazione e datazione di tutte le lezioni.

Questa importantissima nuova fonte di informazioni, tuttavia, non solleva lo studioso dal
porsi alcune fondamentali questioni emerse proprio dal ritrovamento del Documento Moreno.
Stabilita con certezza l'autenticit\'a del Documento, sia per il contenuto (Majorana) che
per l'estensione (Moreno), sorge quindi spontaneo domandarsi se Moreno fosse stato
presente alle lezioni e come abbia ottenuto gli appunti originali di Majorana.
\\
La soluzione pi\'u semplice a tali questioni ci \`e offerta dalla testimonianza di Cesare
Moreno \cite{Moreno}, in base alla quale il padre Eugenio ha sempre concordemente
raccontato ai figli di essere stato presente assiduamente alle lezioni di Majorana e di
aver ricevuto in quella sede gli appunti. Tale testimonianza, tuttavia, si scontra con
quella di altri testimoni del corso, principalmente Sciuti e la Senatore, che invece
affermano di non ricordarsi affatto della presenza di Moreno. Per dirimere la questione,
quindi, sono necessari documenti che non si affidino, possibilmente, solo a ricordi
lontani.

Si vogliono qui ricordare solo alcuni particolari, gi\'a menzionati nelle pagine
precedenti, che, ad avviso di chi scrive, dovrebbero essere tenuti in debito conto per la
risoluzione della ``questione Moreno''.
\\
Innanzitutto \`e da segnalare una coincidenza di date: Moreno rientra dal servizio militare il 30 dicembre 1937, per cui se era sua intenzione seguire qualche corso al suo rientro all'Universit\`a, avrebbe trovato un valido candidato nel corso di Majorana che iniziava proprio a gennaio (contrariamente da altri possibili corsi di suo interesse che gi\'a erano iniziati).
\\
La presenza nel Documento Moreno della numerazione e datazione dettagliata di tutte le
lezioni (comprese le annotazioni particolari per i giorni di festa), non presente negli
originali di Majorana, difficilmente potrebbe essere stata ricostruita, soprattutto molto
tempo dopo la scomparsa di Majorana (vedi pi\'u avanti), da persone non testimoni del
corso, dovendo presumere, inoltre, nessun intento di tipo storico. Di conseguenza,
pur volendo scartare l'ipotesi della presenza di Moreno alle lezioni, si dovrebbe almeno
acconsentire all'ipotesi che Moreno, nello stilare il Documento, abbia avuto a
disposizione almeno {\it due} fonti: sia gli originali di Majorana sia gli appunti o
altre informazioni da testimoni diretti del corso.
\\
Gli appunti originali di Majorana, secondo la testimonianza della Senatore
\cite{Senatore}, \cite{Sena98}, furono consegnati da Majorana stesso alla Senatore il 25
marzo 1938, ed entrarono in possesso di Carrelli (tramite Cennamo) quasi un anno dopo.
Poich\`e la Senatore non ricorda Moreno, quest'ultimo deve aver avuto tra le proprie mani gli
appunti o prima della scomparsa di Majorana (e quindi durante lo svolgimento del corso,
ma non necessariamente a lezione) o circa un anno dopo. La prima ipotesi si scontra con
il fatto che nel Documento Moreno \`e presente anche il testo degli appunti della lezione
del 24 marzo, e gli originali dovevano essere in possesso di Majorana in quella data
(tuttavia tale ipotesi potrebbe anche essere spiegata assumendo che Majorana dava, con
alcuni giorni di anticipo, il testo degli appunti, almeno da un certo punto in avanti).
La seconda ipotesi, al pari, si scontra con la difficolt\'a della ricostruzione storica
dettagliata e particolareggiata delle date di svolgimento delle lezioni.
\\
Dettagli non secondari, emersi dall'analisi anche superficiale del Documento Moreno, sono
poi i seguenti. Moreno lascia circa tre pagine non scritte per ciascuna delle prime
quattro lezioni di Majorana (il cui testo non \`e disponibile). Ci\'o sembrerebbe
indicare, da un lato, che Moreno non ricevette tutte le lezioni di Majorana in un'unica
soluzione o almeno che egli non sapeva che fossero tutte le lezioni (altrimenti non
avrebbe lasciato pagine bianche; inoltre, fino alla lezione N.15 non \`e presente la numerazione
negli originali), mentre dall'altro che Moreno deve aver ricevuto all'inizio almeno pi\'u
di una lezione, in quanto altrimenti non avrebbe potuto desumere la lunghezza tipica di
ciascuna lezione (esattamente circa tre pagine nel proprio formato). Inoltre, il fatto
che la numerazione delle lezioni 1-16 sia stata successivamente cancellata e corretta da
Moreno, secondo quanto riportato in precedenza, potrebbe essere difficilmente spiegabile
se si ammette che Moreno abbia ricevuto tutte le lezioni dopo il corso.
\\
Infine, due ultimi particolari sembrerebbero indicare la presenza di Moreno al corso,
almeno in due occasioni. Il primo \`e la presenza sulla sovraccoperta del Documento
Moreno del riferimento bibliografico del libro di Persico che, secondo quanto ricordato
sopra, fu consigliato da Majorana a Sciuti, il quale \`e l'unico testimone che ha
riportato tale particolare \cite{Sciuti}. Il secondo \`e il fatto che il Documento Moreno
termina con il numero e la data della presunta lezione di sabato 26 marzo, lezione che
purtroppo non fu mai tenuta per la scomparsa di Majorana, e la cui annotazione in un
quaderno \`e quanto meno strana se tale quaderno fosse stato redatto successivamente (e
di molti mesi) la scomparsa.

In conclusione, in base alle conoscenze attuali, sebbene la soluzione pi\'u semplice
testimoniata anche da Cesare Moreno sembri la pi\'u verosimile e attendibile, la
``questione Moreno'' non pu\`o dirsi completamente risolta. Oltre che in tale direzione,
tuttavia, i passi successivi saranno principalmente rivolti allo studio approfondito dei
contenuti scientifici delle sei lezioni inedite, che gi\`a ad un'analisi preliminare si
rivelano particolarmente interessanti.

\section*{Ringraziamenti}

\noindent Voglio qui esprimere la mia profonda gratitudine ad Antonino Drago e a Cesare
Moreno, senza l'intervento dei quali non sarebbe stato possibile rintracciare il
Documento Moreno. Sono inoltre molto grato ad E. Recami per tutto il materiale
documentario e fotografico messomi a disposizione e, per la fattiva collaborazione, a
Francesco Guerra, Bruno Preziosi e a A. De Gregorio, G. Longo, F. Lizzi, E. Majorana jr, G. Mangano, G.
Miele, O. Pisanti.

\vspace{2truecm}


\begin{table}
\begin{center}
\begin{tabular}{|l|l|l|l|l|}
\hline \hline N. & Data & D. M. & D. G. & Argomento
\\ \hline
1. & \bmm Gioved\`i \\ 13-1-38 \eem & mancante & mancante & \bbm
Prolusione. \\ Introduzione: argomenti che saranno trattati nel
corso. \eem \\ \hline
2. & \bmm Sabato \\ 15-1-38 \eem & mancante
& mancante & \\ \hline
3. & \bmm Marted\`i \\ 18-1-38 \eem &
mancante & mancante & \\ \hline
4. & \bmm Gioved\`i \\ 20-1-38
\eem & mancante & mancante & \\ \hline
5. & \bmm Sabato \\ 22-1-38
\eem & mancante & mancante & \\ \hline
6. & \bmm Marted\`i \\ 25-1-38 \eem & presente & presente &
\bbm - Formula della struttura fina. \\ Conferme sperimentali e difficolt\`a. \\ - Il trascinamento del nucleo. \eem \\
\hline
7. & \bmm Gioved\`i \\ 27-1-38 \eem & \underline{presente}
& mancante &
\bbm - Quantizzazione spaziale e propriet\`a magnetiche degli atomi. \\ - Generalit\`a sugli spettri dei metalli alcalini e l'ipotesi dell'elettrone rotante. \eem \\
\hline
8. & \bmm Sabato \\ 29-1-38 \eem & presente & presente &
\bbm - Il principio di Pauli o principio di esclusione e l'interpretazione del sistema periodico degli elementi. \\
- Le condizioni di Sommerfeld per il calcolo dei livelli energetici dei metalli alcalini. \eem \\
\hline
9. & \bmm Gioved\`i \\ 3-2-38 \eem & presente & presente &
\bbm - Lo spettro degli atomi con due elettroni di valenza.
\\ - La teoria classica dell'irraggiamento. \eem \\
\hline
10. & \bmm Sabato \\ 5-2-38 \eem & \underline{presente} & mancante & \bbm - Integrazioni delle Equazioni di Maxwell e applicazioni all'irraggiamento di un sistema oscillante di dimensioni piccole rispetto alla lunghezza d'onda emessa. \\ - Diffusione della luce solare da parte dell'atmosfera. \eem \\
\hline \hline
\end{tabular} \vspace{0.5truecm}
\end{center}
\caption{Ricostruzione dell'indice delle lezioni di Majorana a Napoli, come emerge dal
Documento Moreno. Nelle prime due colonne \`e indicata la numerazione e la data della
lezione. Nelle successive due colonne viene invece indicata la presenza o meno della
lezione nel Documento Moreno (D.M.) e/o nei manoscritti della Domus Galilaeana (D.G.).
Infine nell'ultima colonna viene riportato il titolo dei paragrafi di ciascuna lezione
presenti nel Documento Moreno (o nei manoscritti della D.G.). Una sottolineatura
evidenzia le lezioni non comprese in entrambi gli archivi considerati.}
\end{table}

\setcounter{table}{0}

\begin{table}
\begin{center}
\begin{tabular}{|l|l|l|l|l|}
\hline \hline N. & Data & D. M. & D. G. & Argomento
\\ \hline
11. & \bmm Marted\`i \\ 8-2-38 \eem & \underline{presente} &
mancante &
\bbm - Il principio di relativit\`a nella meccanica classica. \\
- Esperienza di Michelson e Morley. \\
- Le trasformazioni di Lorentz. \eem \\
\hline
12. & \bmm Gioved\`i \\ 10-2-38 \eem & \underline{presente}
& mancante &
\bbm - Il principio di relativit\`a secondo Einstein. \\
- Le leggi di trasformazione dei potenziali elettromagnetici. \eem \\
\hline
13. & \bmm Sabato \\ 12-2-38 \eem & \underline{presente} & mancante & \bbm - Formula di Fresnel ed esperienza di Fizeau. \\
- Invarianza della carica elettrica. \\
- Lo spazio di Minkowski. \\
- Equazioni del movimento per un elettrone in un campo elettromagnetico arbitrario. \eem \\
\hline
14. & \bmm Marted\`i \\ 15-2-38 \eem & \underline{presente}
& mancante &
\bbm - Dinamica relativistica dell'elettrone. \\
- Effetto fotoelettrico. \\
${}$ ~~~ - Relazione fra potenziale e lunghezza d'onda. \\
${}$ ~~~ - Diffusione dell'elettrone libero (formola di Thomson). \eem \\
\hline
15. & \bmm Gioved\`i \\ 17-2-38 \eem & presente & presente
&
\bbm - Effetto Compton. \\
- Esperienza di Franck e Hertz. \eem \\
\hline
16. & \bmm Marted\`i \\ 8-3-38 \eem & presente & presente &
\bbm Nozioni sul calcolo delle matrici: \\
- Spazio di vettori in $n$ dimensioni. \\
- Matrici e operatori lineari. \eem \\
\hline
17. & \bmm Gioved\`i \\ 10-3-38 \eem & presente & presente
&
\bbm - Sistemi unitari. \\
- Operatori Hermitiani. Forme Hermitiane. \eem \\
\hline
18. & \bmm Sabato \\ 12-3-38 \eem & presente & presente &
\bbm - Riduzione contemporanea in forma diagonale di operatori commutabili. \\
- Matrici infinite. \eem \\
\hline \hline
\end{tabular} \vspace{0.5truecm}
\end{center}
\caption{Continuazione.}
\end{table}

\setcounter{table}{0}

\begin{table}
\begin{center}
\begin{tabular}{|l|l|l|l|l|}
\hline \hline N. & Data & D. M. & D. G. & Argomento
\\ \hline
19. & \bmm Gioved\`i \\ 17-3-38 \eem & presente & presente &
\bbm - Integrali di Fourier. \\
La Meccanica Ondulatoria. \\
- Le onde di De Broglie. \eem \\
\hline
20. & \bmm Marted\`i \\ 22-3-38 \eem & presente & presente
&
\bbm - Velocit\`a di fase e velocit\`a di gruppo. \\
- Equazione d'onda non relativistica. Interpretazione statistica dei pacchetti d'onda. \eem \\
\hline
21. & \bmm Gioved\`i \\ 24-3-38 \eem & presente & presente
&
\bbm - Prima estensione dell'interpretazione statistica e relazioni d'incertezza. \eem \\
\hline
? & ? & mancante & \underline{presente} &
\bbm - Sul significato di stato quantico. \\
- Le propriet\`a di simmetria di un sistema nella meccanica classica e quantistica. \\
- Forze di risonanza fra stati non simmetrizzati per perturbazione piccola. Caratteri di simmetria non combinabili. \\
- Consequenze spettroscopiche in atomi con due elettroni. Risonanza fra buche uguali di potenziali e teoria della valenza omeopolare secondo il metodo degli elettroni leganti. \\
- Propriet\`a degli stati simmetrizzati che non si ottengono per perturbazione piccola da stati non simmetrizzati. Bande alternate, idrogeno, ... .
 \eem \\
\hline \hline
\end{tabular} \vspace{0.5truecm}
\end{center}
\caption{Continuazione.}
\end{table}

\end{document}